\documentclass[12pt]{article}
\usepackage{amsmath}
\usepackage[dvips]{graphicx}
\input amssym

\begin{document}

\def\prg#1{\medskip\noindent{\bf #1}}  \def\ra{\rightarrow}
\def\lra{\leftrightarrow}              \def\Ra{\Rightarrow}
\def\nin{\noindent}                    \def\pd{\partial}
\def\dis{\displaystyle}                \def\inn{\,\hook\,}
\def\grl{{GR$_\Lambda$}}               \def\Lra{{\Leftrightarrow}}
\def\cs{{\scriptstyle\rm CS}}          \def\ads3{{\rm AdS$_3$}}
\def\Leff{\hbox{$\mit\L_{\hspace{.6pt}\rm eff}\,$}}
\def\bull{\raise.25ex\hbox{\vrule height.8ex width.8ex}}
\def\ric{{Ric}}                      \def\tric{{(\widetilde{Ric})}}
\def\Lie{{\cal L}\hspace{-.7em}\raise.25ex\hbox{--}\hspace{.2em}}
\def\sS{\hspace{2pt}S\hspace{-0.83em}\diagup}   \def\hd{{^\star}}
\def\dis{\displaystyle}                 \def\ul#1{\underline{#1}}
\def\nb#1{\marginpar{\large #1}}

\def\hook{\hbox{\vrule height0pt width4pt depth0.3pt
\vrule height7pt width0.3pt depth0.3pt
\vrule height0pt width2pt depth0pt}\hspace{0.8pt}}
\def\semidirect{\;{\rlap{$\supset$}\times}\;}
\def\first{\rm (1ST)}       \def\second{\hspace{-1cm}\rm (2ND)}
\def\bm#1{\hbox{{\boldmath $#1$}}}

\def\G{\Gamma}        \def\S{\Sigma}        \def\L{{\mit\Lambda}}
\def\D{\Delta}        \def\Th{\Theta}
\def\a{\alpha}        \def\b{\beta}         \def\g{\gamma}
\def\d{\delta}        \def\m{\mu}           \def\n{\nu}
\def\th{\theta}       \def\k{\kappa}        \def\l{\lambda}
\def\vphi{\varphi}    \def\ve{\varepsilon}  \def\p{\pi}
\def\r{\rho}          \def\Om{\Omega}       \def\om{\omega}
\def\s{\sigma}        \def\t{\tau}          \def\eps{\epsilon}
\def\nab{\nabla}      \def\btz{{\rm BTZ}}   \def\heps{\hat\eps}
\def\bu{{\bar u}}     \def\bv{{\bar v}}     \def\bs{{\bar s}}
\def\bx{{\bar x}}     \def\by{{\bar y}}
\def\tphi{{\tilde\vphi}}  \def\tt{{\tilde t}}
\def\bara{{\bar\a}}     \def\bp{{\bar p}}     \def\bq{{\bar q}}

\def\tG{{\tilde G}}   \def\cF{{\cal F}}      \def\bH{{\bar H}}
\def\cL{{\cal L}}     \def\cM{{\cal M }}     \def\cE{{\cal E}}
\def\cH{{\cal H}}     \def\hcH{\hat{\cH}}    \def\hS{{\hat S}}
\def\cK{{\cal K}}     \def\hcK{\hat{\cK}}    \def\cT{{\cal T}}
\def\cO{{\cal O}}     \def\hcO{\hat{\cal O}} \def\cV{{\cal V}}
\def\tom{{\tilde\omega}}                     \def\cE{{\cal E}}
\def\cR{{\cal R}}    \def\hR{{\hat R}{}}     \def\hL{{\hat\L}}
\def\tE{{\tilde E}}  \def\tH{{\tilde H}}     \def\tX{{\tilde X}}
\def\tW{{\tilde W}}  \def\tR{{\widetilde R}} \def\tcL{{\tilde\cL}}
\def\hy{{\hat y}\hspace{1pt}}  \def\tcO{{\tilde\cO}}
\def\tric{{\widetilde\ric}}

\def\nn{\nonumber}                    \def\vsm{\vspace{-7pt}}
\def\be{\begin{equation}}             \def\ee{\end{equation}}
\def\ba#1{\begin{array}{#1}}          \def\ea{\end{array}}
\def\bea{\begin{eqnarray} }           \def\eea{\end{eqnarray} }
\def\beann{\begin{eqnarray*} }        \def\eeann{\end{eqnarray*} }
\def\beal{\begin{eqalign}}            \def\eeal{\end{eqalign}}
\def\lab#1{\label{eq:#1}}             \def\eq#1{(\ref{eq:#1})}
\def\bsubeq{\begin{subequations}}     \def\esubeq{\end{subequations}}
\def\bitem{\begin{itemize}}           \def\eitem{\end{itemize}}
\renewcommand{\theequation}{\thesection.\arabic{equation}}

\title{Siklos waves in Poincar\'e gauge theory}

\author{M. Blagojevi\'c and B. Cvetkovi\'c\footnote{
        Email addresses: {mb@ipb.ac.rs,
                          cbranislav@ipb.ac.rs}}\\
Institute of Physics, University of Belgrade \\
                      Pregrevica 118, 11080 Belgrade, Serbia}
\date{\today}
\maketitle

\begin{abstract}
A class of Siklos waves, representing exact vacuum solutions of general
relativity with a cosmological constant, is extended to a new class of
Siklos waves with torsion, defined in the framework of the Poincar\'e
gauge theory. Three particular exact vacuum solutions of this type, the
generalized Kaigorodov, the homogeneous and the exponential solution, are
explicitly constructed.
\end{abstract}
\section{Introduction}
\setcounter{equation}{0}

The first complete formulation of the idea of (internal) gauge invariance
was given in Weyl's classic paper \cite{x1}. A significant progress in
this direction has been achieved somewhat later by Yang, Mills and Utiyama
\cite{x2,x3}. It opened a new perspective for understanding gravity as a
gauge theory, the perspective that was realized by Kibble and Sciama
\cite{x4} in their proposal of a new theory of gravity, known as the
Poincar\'e gauge theory (PGT). The PGT is a gauge theory of the Poincar\'e
group, with an underlying Riemann-Cartan (RC) geometry of spacetime
\cite{x5,x6}. In this approach, basic gravitational variables are the
tetrad field $b^i$ and the Lorentz connection $\om^{ij}$ (1-forms), and
the related field strengths are the torsion $T^i=db^i+\om^i{_m}\wedge
b^{mj}$ and the curvature $R^{ij}=d\om^{ij}+\om^i{_m}\wedge\om^{mj}$
(2-forms). At a more physical level, the source of gravity in PGT is
matter possessing both the energy-momentum and spin currents. The
importance of the Poincar\'e symmetry in particle physics leads one to
consider PGT as a favorable framework for describing the gravitational
phenomena.

Based on the experience stemming from Einstein's general relativity, it is
known that exact solutions play a crucial role in developing our
understanding of the geometric and physical content of a gravitational
theory; for a review, see Refs. \cite{x7,x8,x9,x10}. An important set of
these solutions refers to exact gravitational waves, the structure of
which has been studied also in PGT \cite{x11}. In the present work, we
focus on a particular class of the gravitational waves, the class of
Siklos waves that are vacuum solutions of general relativity with a
cosmological constant (\grl), propagating on the AdS background
\cite{x12}. By generalizing the ideas developed in three dimensions
\cite{x13}, we construct here a class of the four-dimensional Siklos waves
with torsion as vacuum solutions of PGT.

The paper is organized as follows. In section 2, we give a short account
of the Siklos waves in the tetrad formulation of \grl. In section 3, we
show that Siklos waves are torsion-free vacuum solutions of PGT. In
section 4, we introduce new vacuum solutions of PGT, the Siklos waves with
torsion, by modifying the Siklos geometry in a manner that preserves the
radiation nature of the original configuration. That is achieved by an
ansatz for the RC connection that produces only the tensorial irreducible
mode of the torsion with $J^P=2^+$. The PGT field equations are simplified
and shown to depend only on three parameters, including the mass of the
torsion mode. In sections 5, 6 and 7, we describe three different vacuum
solutions belonging to the class of Siklos waves with torsion: the
generalized Kaigorodov, the homogeneous and the exponential solution.
section 7 is devoted to concluding remarks, and two appendices contain
some technical details.

Our conventions are as follows. We use the Poincar\'e coordinates
$x^\m=(u,v,x,y)$ as the local coordinates; the Latin indices $(i,j,...)$
refer to the local Lorentz (co)frame and run over $(+,-,2,3)$, $b^i$ is
the tetrad (1-form), $h_i$ is the dual basis (frame), such that $h_i\inn
b^k=\d_i^k$; the volume 4-form is $\heps=b^+\wedge b^-\wedge b^2\wedge
b^3$, the Hodge dual of a form $\a$ is $\hd\a$, with $\hd 1=\heps$,
totally antisymmetric tensor is defined by $\hd(b_i\wedge b_j\wedge
b_k\wedge b_m)=\ve_{ijkm}$ and normalized to $\ve_{+-23}=1$; in the rest
of the paper, the exterior product of forms is implicit.

\section{Siklos waves in GR\bm{_\L}}\label{sec2}
\setcounter{equation}{0}

Siklos waves were introduced as a class of exact gravitational waves
propagating on the AdS background \cite{x12}. In the Poincar\'e coordinates
$x^\m=(u,v,x,y)$, the Siklos metric is given by
\be
ds^2=\frac{\ell^2}{y^2}\left[2du(Hdu+dv)-dx^2-dy^2\right]\, ,   \lab{2.1}
\ee
with $H=H(u,x,y)$. It admits the null Killing vector field $\pd_v$ that is
not covariantly constant, the wave fronts are surfaces of constant $u$ and
$v$, and the case $H=0$ corresponds to the AdS background. The metric
\eq{2.1} coincides with a special subclass of the Kundt class
\cite{x9,x10}, and is obviously conformal to pp-waves. The physical
interpretation of the Siklos waves was investigated by Podolsk\'y
\cite{x14,x15}.

Now, we give a short account of the Siklos waves in the tetrad formulation
of \grl, which allows for a simpler generalization to PGT. First, we
choose the tetrad field in the form
\be
b^+:=\frac{\ell}{y}du\, ,\qquad
b^-:=\frac{\ell}{y}(Hdu+dv)\, ,\qquad
b^2:=\frac{\ell}{y}dx\, ,\qquad
b^3:=\frac{\ell}{y}dy\, ,                                       \lab{2.2}
\ee
so that the line element becomes
$ds^2=2b^+b^--(b^2)^2-(b^3)^2\equiv\eta_{ij}b^ib^j$, where $\eta$ is the
half-null Minkowski metric,
$$
\eta_{ij}=\left( \ba{cccc}
             0 & 1 & 0 & 0  \\
             1 & 0 & 0 & 0  \\
             0 & 0 & -1& 0  \\
             0 & 0 & 0 & -1
                \ea
       \right)\, .
$$
The dual frame $h_i$ is given by
\be
h_+=\frac{y}{\ell}(\pd_u-H\pd_v)\,,\qquad
h_-=\frac{y}{\ell}\pd_v\, ,\qquad h_2=\frac{y}{\ell}\pd_x\, ,\qquad
h_3=\frac{y}{\ell}\pd_y\, .
\ee

Next, we introduce the Riemannian connection $\om^{ij}$ by imposing the
condition of vanishing torsion, $\nab b^i:=db^i+\om^i{_m}b^m =0$, which
yields
\bsubeq\lab{2.4}
\bea
&&\om^{+-},\om^{+2}=0\, ,\qquad\om^{+3}=\frac{1}{\ell}b^+\, ,
  \qquad  \om^{23}=\frac{1}{\ell}b^2\, ,                        \nn\\
&&\om^{-2}=-\frac{y}{\ell}\pd_x Hb^+\, ,\qquad
  \om^{-3}=\frac{1}{\ell}b^- -\frac{y}{\ell}\pd_y H b^+\, .
\eea
The wave nature of the Siklos wave is clearly seen by rewriting
$\om^{ij}$ in the form
\be
\om^{ij}=\bar\om^{ij}+k^i(h^j\inn H)b^+\, ,
\ee
\esubeq
where $\bar\om^{ij}=\om^{ij}(H=0)$ refers to the AdS background, and the
second term is the radiation piece, characterized by the null vector
$k^i=(k^+,k^-,k^2,k^3)=(0,1,0,0)$.

Now, one can calculate the Riemannian curvature:
\bea
&&R^{+j}=\frac{1}{\ell^2}b^+b^j\, ,\qquad
  R^{23}=\frac{1}{\ell^2}b^2b^3\, ,                             \nn\\
&&R^{-2}=\frac{1}{\ell^2}b^-b^2
  +\frac{1}{\ell^2}(y^2\pd_{xx}H-y\pd_yH)b^+b^2
  +\frac{1}{\ell^2}(y^2\pd_{xy}H)b^+b^3\, ,                     \nn\\
&&R^{-3}=\frac{1}{\ell^2}b^-b^3
  +\frac{1}{\ell^2}(y^2\pd_{yy}H-y\pd_yH)b^+b^3
  +\frac{1}{\ell^2}(y^2\pd_{xy}H)b^+b^2\, ,                     \lab{2.5}
\eea
where we use $\pd_{xx}:=\pd^2/\pd x^2$ etc. The Ricci curvature
$\ric^i=h_m\hook R^{mi}$ and the scalar curvature $R=h_i\hook\ric^i$ are
found to be
\bea
&&\ric^m=\frac{3}{\ell^2}b^m\, ,\qquad m=+,2,3,                 \nn\\
&&\ric^-=\frac{3}{\ell^2}b^-
  +\frac{1}{\ell^2}(y^2\pd_{xx}H+y^2\pd_{yy}H-2y\pd_yH)b^+\, ,  \nn\\
&&R=\frac{12}{\ell^2}\, .
\eea

Dynamical structure of \grl\ is defined by the action $I_\L=-\int d^4
x\sqrt{-g}\, (a_0R+2\L)$. The corresponding \emph{vacuum} field equations
can be suitably written in the traceless form as
\be
\ric^i-\frac{1}{4}Rb^i=0\, .                                    \lab{2.7}
\ee
As a consequence, the metric function $H$ must obey
\be
y^2\left(\pd_{xx}H+\pd_{yy}H\right)-2y\pd_y H=0\, .             \lab{2.8}
\ee
The profile ($u$-dependence) of the Siklos wave may be arbitrary.

We display here three special solutions of \eq{2.8} discussed by Siklos
\cite{x12}:
\bea
&&H_1=y^3,\, {\rm Kaigorodov's~ solution~ (1963)};              \nn\\
&&H_2=\arctan(x/y)+xy/(x^2+y^2)\,,\qquad\tilde H_2=(x^2+y^2)H_2;\nn\\
&&H_3=C_1e^x(\cos y+y\sin y)+C_2e^x(\sin y-y\cos y)\, .         \nn
\eea
Note that Defrise's metric (1969), with $H=1/y^2$, is \emph{not} a vacuum
solution of \grl\ \cite{x15}.

\section{Siklos waves as torsion-free solutions of PGT}
\setcounter{equation}{0}

In this section, we show that the Siklos spacetime of the previous section
is an exact Riemannian solution of PGT in vacuum.

Starting from the general PGT dynamics described in Appendix B, one can
easily derive its reduced form in the Riemannian sector of PGT,
characterized by $T^i=0$. First, we note that the only nonvanishing
irreducible components of the Riemannian curvature are ${}^{(1)}R^{ij}$,
${}^{(4)}R^{ij}$ and ${}^{(6)}R^{ij}$, defined in Appendix A. And second,
the condition $T^i=0$ implies that dynamical evolution of the Riemannian
solutions in PGT is described by a reduced form of the general field
equations \eq{B.3}:
\bsubeq\lab{3.1}
\bea
(1ST)&&E_i=0\, ,                                                \nn\\
(2ND)&&\nab H_{ij}=0\, .                                        \lab{3.1a}
\eea
Here, the Riemannian expressions for $E_i$ and $H_{ij}$ are obtained
directly from the corresponding PGT formulas (see Appendix B) in the limit
$T^i=0$:
\bea
&&H_{ij}=-2a_0\hd(b^ib^j)
  +2\hd(b_1{}^{(1)}R_{ij}+b_4{}^{(4)}R_{ij}+b_6{}^{(6)}R_{ij})\,,\nn\\
&&E_i:=h_i\inn L_G -\frac{1}{2}(h_i\inn R^{mn})H_{mn}\, .
\eea
\esubeq

As shown in Ref. \cite{x5}, the field equations \eq{3.1} are satisfied for
any configuration in which the traceless symmetric Ricci tensor vanishes:
\be
\ric_{(ij)}-\frac{1}{4}\eta_{ij}R=0\, .                         \lab{3.2}
\ee
Comparing this result with the \grl\ field equation \eq{2.7}, one
concludes that any vacuum solution of \grl\ is automatically a
torsion-free solution of PGT. In particular, this is true for the Siklos
metric.

It is useful to explore this general statement in details. Using the
geometry of the Siklos spacetime found in the previous section, the
content of Eqs. \eq{3.1a} is found to be:
\bea
(1ST)&&\bigl(b_4+b_6-a_0\ell^2\bigr)
     y\Bigl[y(\pd_{xx}H+\pd_{yy}H)-2\pd_yH\Bigr]=0\, ,          \nn\\
     && 3a_0+\ell^2\L=0,                                        \nn\\
(2ND)&&(b_1+b_4)y^2\pd_x
               \Bigl[y(\pd_{xx}H+\pd_{yy}H)-2\pd_{y}H\Bigr]=0\,,\nn\\
     &&(b_1+b_4)y^2\pd_y
               \Bigl[y(\pd_{xx}H+\pd_{yy}H)-2\pd_{y}H\Bigr]=0\,.
\eea
For the generic values of the Lagrangian parameters $a_0,b_1,b_4,b_6$,
dynamical content of these equations is obviously the same as in \grl,
since the metric function $H$ must be such that
\be
\hS H:=y\left(\pd_{xx}H+\pd_{yy}H\right)-2\pd_y H=0\, .         \lab{3.4}
\ee
Thus, although PGT has a rather different dynamical structure as compared
to \grl, the class of Riemannian Siklos spacetimes is still an exact
vacuum solution of PGT.

\section{Siklos waves with torsion}\label{sec4}
\setcounter{equation}{0}

We are now ready to generalize the previous results by constructing a new,
non-Riemannian class of Siklos waves, the Siklos waves with torsion.

\subsection{Geometry of the ansatz}

We wish to introduce torsion in a manner that \emph{preserves the
radiation nature} of the Riemannian Siklos waves of \grl, relying on the
approach proposed in \cite{x13}.

We start the construction by assuming that the tetrad field in PGT retains
its Riemannian form \eq{2.2}. Then, by noting that the radiation piece of
the Riemannin connection \eq{2.4} has the form
$(\om^{ij})^R=k^i(h^{j\m}\pd_\m H)b^+$, we assume that the new RC
connection is given by
\bsubeq
\be
\om^{ij}=\bar\om^{ij}+k^ih^{j\m}(\pd_\m H+K_\m)b^+\, ,
\ee
where the form of $K_\m$ is defined by
\bea
&&K_\m=(0,0,K_x,K_y)\, ,                                        \nn\\
&&K_x=K_x(u,x,y)\,,\qquad K_y=K_y(u,x,y)\, .
\eea
\esubeq
This ansatz modifies only two components of the Riemannian connection
\eq{2.4}:
$$
\om^{-2}=-\frac{y}{\ell}(\pd_x H+K_x)b^+\, ,\qquad
\om^{-3}=\frac{1}{\ell}b^- -\frac{y}{\ell}(\pd_y H+K_y)b^+\, .
$$
The new terms in the connection are related to the torsion of spacetime:
\be
T^-=\frac{y}{\ell}(K_xb^+b^2+K_y b^+b^3)\, ,\qquad T^+,T^2,T^3=0\,.
\ee
The only nonvanishing irreducible torsion piece is the tensor piece
${}^{(1)}T^i$, with ${}^{(1)}T^-=T^-$.

Denoting the Riemannian curvature found in section 2 by $\tR^{ij}$, the
new, RC curvature is found to have the form:
\bsubeq\lab{4.3}
\bea
&&R^{+j}=\frac{1}{\ell^2}b^+b^j\, ,\qquad
  R^{23}=\frac{1}{\ell^2}b^2b^3\, ,                             \nn\\
&&R^{-2}=\tR^{-2} +\frac{1}{\ell^2}(y^2\pd_x K_x-yK_y)b^+b^2
         +\frac{1}{\ell^2}(y^2\pd_yK_x)b^+b^3\, ,               \nn\\
&&R^{-3}=\tR^{-3} +\frac{1}{\ell^2}(y^2\pd_yK_y)b^+b^3
         +\frac{1}{\ell^2}(y^2\pd_xK_y+yK_x)b^+b^2\, .
\eea
Note that the radiation piece of $R^{ij}$ is proportional to the null
vector $k^i=(0,1,0,0)$. The corresponding Ricci and scalar curvatures are:
\bea
&&\ric^m=\frac{3}{\ell^2}b^a\, ,\quad m=+,2,3,                  \nn\\
&&\ric^-=\tric{}^-
  +\frac{1}{\ell^2}(y^2\pd_xK_x+y^2\pd_yK_y-yK_y)b^+\, ,        \nn\\
&&R=\frac{12}{\ell^2}\, .
\eea
\esubeq
The nonvanishing irreducible components of the curvature are
${}^{(n)}R^{ij}$ for $n=1,4,6$ (as in \grl) and $n=2$. Quadratic
invariants of the field strengths are regular:
$$
R^{ij}\hd R_{ij}=\frac{12}{\ell^4}\,\heps\, ,\qquad T^i\hd T_i=0\, .
$$

\subsection{Field equations}

Dynamical content of our ansatz is effectively described by the RC
Lagrangian \eq{B.1} with nonvanishing parameters
$(a_0,\L;a_1,b_1,b_2,b_4,b_6)$, and the associated PGT field equations
\eq{B.3}. Explicit calculation of the 2nd field equation in \eq{B.3},
denoted shortly by $\cF^{ij}$, is shown to have two nontrivial components,
$\cF^{-2}$ and $\cF^{-3}$. After introducing the quantity $\hS H$ as in
Eq. \eq{3.4}, these components take the respective forms:
\bsubeq\lab{4.4}
\bea
&&b_1\left(y\pd_x \hS H +y^2\pd_{xx}K_x+y^2\pd_{yy}K_x
     -2y\pd_x K_y\right)                                        \nn\\
&&+b_2(y^2\pd_{yy}K_x-y^2\pd_{xy}K_y-y\pd_xK_y)                 \nn\\
&&+b_4\left(y\pd_x\hS H
     +y^2\pd_{xx}K_x+y^2\pd_{xy}K_y-y\pd_x K_y\right)           \nn\\
&&+2(b_6-b_1+a_1\ell^2-a_0\ell^2)K_x=0\, ,                      \lab{4.4a}
\eea
and
\bea
&&b_1\left(y\pd_y\hS H
     +y^2\pd_{xx}K_y+y^2\pd_{yy}K_y+2y\pd_x K_x\right)          \nn\\
&&+b_2(-y^2\pd_{xy}K_x+y^2\pd_{xx}K_y+y\pd_x K_x)               \nn\\
&&+b_4\left(y\pd_y\hS H
     +y^2\pd_{xy}K_x+y^2\pd_{yy}K_y+y\pd_xK_x\right)            \nn\\
&&+2(b_6-b_1+a_1\ell^2-a_0\ell^2)K_y=0\, .                      \lab{4.4b}
\eea
The content of the 1st field equation is much simpler. To have the smooth
limit for vanishing torsion, we require $3a_0+\ell^2\L=0$, whereupon the
1st equation reads
\bea
&&(b_4+b_6-a_0\ell^2)\hS H                                      \nn\\
&&+(b_4+b_6-a_0\ell^2+a_1\ell^2)(y\pd_xK_x+y\pd_yK_y-K_y)=0\, . \lab{4.4c}
\eea
\esubeq

The form of the differential equations \eq{4.4} appears to be rather
complicated \cite{x16}. However, there exists a suitable reformulation
that makes their content much more transparent. To see that, we first
rewrite Eq. \eq{4.4c} in the form
\bsubeq\lab{4.5}
$$
\hS H=\s(y\pd_xK_x+y\pd_yK_y-K_y)\, ,\qquad
\s:=-\left(1+\frac{a_1\ell^2}{b_4+b_6-a_0\ell^2}\right)\, .   \eqno(4.5c)
$$
Then, by substituting the expressions for $y\pd_x\hS H$ and $y\pd_y\hS H$
into \eq{4.4a} and \eq{4.4b}, and dividing the resulting equations by
$(b_1+b_4)(\s+1)$, one obtains
\bea
&&(y^2\pd_{xx} +\r y^2\pd_{yy}+2\ell^2\m^2)K_x
       +\bigl[(1-\r)y^2\pd_{xy}-(1+\r)y\pd_x\bigr]K_y=0\, ,       \\
&&(y^2\pd_{yy} +\r y^2\pd_{xx}+2\ell^2\m^2)K_y
       +\bigl[(1-\r)y^2\pd_{xy}+(1+\r)y\pd_x\bigr]K_x=0\, .
\eea
\esubeq
where
\be
\r:=\frac{b_1+b_2}{(b_1+b_4)(\s+1)}\, ,\qquad
  \m^2:=\frac{a_1-a_0+(b_6-b_1)/\ell^2}{(b_1+b_4)(\s+1)}\, . \nn
\ee
The final equations \eq{4.4} contain only three independent parameters,
$\s,\r$ and $\m^2$, which makes it much easier to find some specific
solutions for the Siklos waves with torsion.

The parameter $\m^2$ has a simple physical interpretation. As the
linearized PGT analysis shows, possible torsion excitations around the
Minkowski background are modes with spin-parity $J^P=0^\pm,1^\pm,2^\pm$
\cite{x18}. In particular, the spin-$2^+$ state is associated to the
tensorial piece of the torsion, and its mass is
$$
\bar\m^2=\frac{a_0(a_1-a_0)}{a_1(b_1+b_4)}\, .
$$
For $1/\ell^2\to 0$, the coefficient $\m^2$ tends exactly to $\bar\m^2$,
whereas for finite (and positive) $\ell^2$, $\m^2$ is associated to the
spin-$2^+$ torsion excitation with respect to the AdS background.

In what follows, we will present three exact solutions of the PGT field
equations \eq{4.4}, enlightening thereby basic dynamical aspects of the
Siklos waves with torsion. All the integration ``constants" appearing in
these solutions are functions of $u$.

\section{Kaigorodov-like solution}
\setcounter{equation}{0}

Motivated by the form of the Kaigorodov solution of \grl\ (section
\ref{sec2}), we consider now a class of PGT configurations for which the
functions $H,K_x$ and $K_y$ are $x$-independent. Then, the field equations
\eq{4.5} take a much simpler form:
\bsubeq\lab{5.1}
\bea
&&(\r y^2\pd_{yy}+2\m^2\ell^2)K_x=0\,,                          \lab{5.1a}\\
&&(y^2\pd_{yy}+2\m^2\ell^2)K_y=0\, ,                            \lab{5.1b}\\
&&y\pd_{yy}H-2\pd_y H=\s(y\pd_y-1)K_y\, .                       \lab{5.1c}
\eea
\esubeq

The Euler--Fuchs differential equation \eq{5.1a} is solved by the ansatz
$K_x=y^\a$, where $\a$ is restricted by the requirement
$\a^2-\a+2\m^2\ell^2/\r=0$, which implies
\be
\a_\pm=\frac{1}{2}\pm p\, ,\qquad p:=\frac{1}{2}\sqrt{1-8\m^2\ell^2/\r}\, .
\ee
(a1) For $8\m^2\ell^2/\r<1$ (real $p$):
\bsubeq\lab{5.3}
\be
K_x=\sqrt{y}\left(A_1y^p+A_2y^{-p}\right)\, .
\ee
(a2) For $8\m^2\ell^2/\r>1$ (imaginary $p$, $q:=|p|$):
\be
K_x=\sqrt{y}\bigl[A_3\cos(q\ln y) +A_4\sin(q\ln y)\bigr]\, .
\ee
(a3) For $8\m^2\ell^2/\r=1$ ($p=0$):
\be
K_x=\sqrt{y}\left(A_5+A_6\ln y\right)\,.
\ee
\esubeq

Equation \eq{5.1b} follows from \eq{5.1a} in the limit $\r\to 1$. Hence,
using the notation
\be
\bara_\pm=\frac{1}{2}\pm \bp\, ,\qquad \bp:=\frac{1}{2}\sqrt{1-8\m^2\ell^2}\, ,
\qquad \bq=|\bp|\,,
\ee
the solutions for $K_y$ can be obtained from Eqs. \eq{5.3} by the replacements
$p\to\bp$, $q\to\bq$.\\
(b1) For $8\m^2\ell^2<1$:
\bsubeq\lab{5.5}
\be
K_y=\sqrt{y}\left(B_1y^\bp+B_2y^{-\bp}\right)\, .
\ee
(b2) For $8\m^2\ell^2>1$:
\be
K_y=\sqrt{y}\bigl[B_3\cos(\bq\ln y) +B_4\sin(\bq\ln y)\bigr]\, ,
\ee
(b3) For $8\m^2\ell^2=1$:
\be
K_y=\sqrt{y}\left(B_5+B_6\ln y\right)\,.
\ee
\esubeq

Knowing the form of $K_y$, one can integrate Eq. \eq{5.1c} to obtain the
metric function $H$. Let us first find a particular solution
$H_{(i)}$ of the inhomogeneous equation \eq{5.1c}.\\
(c1) For $8\m^2\ell^2<1$:
\bsubeq\lab{5.6}
\be
H_{(i)}=\s y^{3/2}\left(\frac{(\bara_+-1)}{(\bara_++1)(\bara_+-2)}B_1y^\bp
         +\frac{(\bara_--1)}{(\bara_-+1)(\bara_--2)}B_2y^{-\bp}\right)\,.
                                                                \lab{5.6a}
\ee
(c2) For $8\m^2\ell^2>1$:
\be
H_{(i)}=\frac{2\s}{9+4\bq^2}\,y^{3/2}\bigl[
        (B_3-2B_4\bq)\cos(\bq\ln y) +(B_4+2B_3\bq)\sin(\bq\ln y)\bigr]\,.
\ee
(c3) For $8\m^2\ell^2=1$:
\be
H_{(i)}=\frac{2\s}{9}\,y^{3/2}\left(B_5-2B_6+B_6\ln y\right)\, .
\ee
\esubeq
Adding to $H_{(i)}$ the solution of the homogeneous equation \eq{5.1c},
that is the Kaigorodov solution $H_1$ from section \ref{sec2}, one obtains
the complete solution:
\be
H=H_1+H_{(i)}\, ,\qquad H_1=Dy^3\, .
\ee
Thus, the existence of torsion has a direct influence on the form of
metric.

The above solutions for $K_x,K_y$ and $H$ define a Kaigorodov wave with
torsion as a vacuum solution of PGT.

\subsection*{Asymptotic AdS limit}

It is interesting to note that the Kaigorodov solution in \grl\ is
asymptotically AdS, as follows from the asymptotic relation $H=\cO(y^3)$
for $y\to 0$, and the form of the Riemannian curvature \eq{2.5}. In PGT,
the presence of torsion makes the situation not so simple. Namely, the
condition that the RC curvature $R^{ij}$ in \eq{4.3} has the AdS
asymptotics produces two types of requirements: the first one is obtained
from the non-Riemannian piece of $R^{ij}$,
\bsubeq\lab{5.8}
\bea
&&yK_x\to 0\, ,\qquad yK_y\to 0\, ,                             \lab{5.8a}\\
&& y^2\pd_y K_x\to 0\, ,\qquad y^2\pd_y K_y\to 0\, ,
\eea
and the second from the Riemannian piece:
\be
y\pd_y H_{(i)}\to 0\, ,\qquad y^2\pd_{yy} H_{(i)} \to 0\,.
\ee
\esubeq
Further analysis goes as follows.

(i) In the sector with $8\m^2\ell^2/\r\ge 1$ and $8\m^2\ell^2\ge 1$, one
can directly verify that the solutions for $K_x,K_y$ and $H_{(i)}$ satisfy
the requirements \eq{5.8}.

(ii) In the complementary sector with $8\m^2\ell^2/\r<1$ and
$8\m^2\ell^2<1$, one finds that the requirements \eq{5.8} are valid for
$p<1$ and $\bp<1$, or equivalently, for
\be
8\m^2\ell^2/\r>-1\quad{\rm and}\quad 8\m^2\ell^2>-1\, .         \lab{5.9}
\ee

Continuing with exploring the asymptotic properties of the torsion, we see
that \eq{5.8a} implies $T^i\to 0$ for $y\to 0$. Thus, the choice of
parameters described in \eq{5.9} ensures that the Kaigorodov-like solution
has an AdS asymptotic behavior, with vanishing torsion. Clearly, in the
physical sector with $\m^2\ge 0$, the second condition in \eq{5.9} is
automatically satisfied.

\subsection*{Defrise-like solution as special case}

It is interesting to observe that the form of $H_{(i)}$ in \eq{5.6a}
allows us to obtain a generalized Defrise solution, defined in section
\ref{sec2}, as a special case of the Kaigorodov wave with torsion. Namely,
by choosing $D=0$ one eliminates $H_1$ from $H$, whereupon the term
$H_{(i)}$, specified by $B_1=0$ and $\bp=7/2$, becomes identical to the
Defrise metric function:
\be
H=H_{(i)}\sim 1/y^2\,.
\ee
The restriction $\bp=7/2$ refers to the tachyonic sector of the $2^+$
torsion mode, with $\m^2\ell^2=-6$. The above result for $H$, combined
with the corresponding expressions for $K_x$ and $K_y$, defines the
Defrise solution with torsion as a \emph{vacuum} solution of PGT. In
contrast to that, the corresponding solution in \grl\ exists only in the
presence of \emph{matter}. One should stress that the metric function $H$
originates purely from the torsional term $H_{(i)}$.

\section{Homogeneous solution}
\setcounter{equation}{0}

Let us now look for a solution in which $K_x,K_y,H$ are homogeneous
functions of $y$ and $x$:
$$
K_x=f_x(t)\, ,\qquad K_y=f_y(t)\, ,\qquad H=h(t)\, ,\qquad t:=y/x\,.
$$
As a consequence, the field equations \eq{4.5} become:
\bsubeq\lab{6.1}
\bea
&&(t^4+\r t^2)f_x''+2t^3f_x'+2\m^2f_x-(1-\r)t^3f_y''
                                     +2\r t^2f_y'=0\, ,         \lab{6.1a}\\
&&(t^2+\r t^4)f_y''+2\r t^3f_y'+2\m^2f_y-(1-\r)t^3f_x''
                                        -2t^2f_x'=0\, ,         \lab{6.1b}\\
&&\hS H=\s(-t^2f'_x+tf'_y-f_y)\, ,                              \lab{6.1c}
\eea
\esubeq
where $\hS H=y[2t(t^2-1)h'+(t^4+t^2)h'']$.

The set of equations \eq{6.1} represents a system of ordinary,
second-order, linear differential equations. The system is significantly
simplified by assuming that the metric function $H$ retains the same form
as in \grl, so that $\hS H=0$. Consequently, the right hand side of Eq.
\eq{6.1c} vanishes, $-t^2f'_x+tf'_y-f_y=0$, which implies
\be
f_x=\frac{1}{t}f_y+B\, ,                                        \lab{6.2}
\ee
where $B=B(u)$. Substituting this expression into \eq{6.1a} and \eq{6.1b},
one obtains
\bsubeq\lab{6.3}
\bea
&&\r t^2(t^2+1)f_y''+2\r t(t^2-1)f_y'+2(\r+\mu^2\ell^2)f_y+2\mu^2 tB=0\,,\\
&&\r t^2(t^2+1)f_y''+2\r t(t^2-1)f_y'+2(\r+\mu^2\ell^2)f_y=0\,.
\eea
\esubeq
Taking the difference of these two equations yields
$$
\mu^2 B=0\, .
$$
Hence, either $\mu^2$ or $B$ has to vanish.

\subsection*{Case \bm{\mu^2=0}}

Assuming $\r\ne 0$, the set of equations \eq{6.3} reduces to
$$
t^2(t^2+1)f_y''+2t(t^2-1)f_y'+2f_y=0\, .
$$
Hence, the general solution for $f_y$ is given by
\be
f_y=C_1\frac t{t^2+1}+C_2\frac{t^2}{t^2+1}\, ,
\ee
$f_x$ is determined by \eq{6.2}, and the metric function has the same form
as in \grl:
\be
h=C_3\left (-\arctan t+\frac{t}{1 + t^2}\right) +C_4\, .        \lab{6.5}
\ee
As before, all the integration constants are functions of $u$.

\subsection*{Case \bm{B=0}}

In this case, the set of equations \eq{6.3} reduces to
$$
t^2(t^2+1)f_y''+2t(t^2-1)f_y'
               +2\left(1+\frac{\mu^2\ell^2}{\r}\right)f_y=0\, .
$$
(d1) For $8\m^2\ell^2/\r\ne 1$:
\bsubeq\lab{6.6}
\bea
\hspace{-9pt}
f_y=C_5t^{\frac{3}{2}-\xi}\,
   {}_2F_1\left(\frac{3}{4}-\frac{\xi}{2},
                \frac{5}{4}-\frac{\xi}{2};1-\xi;-t^2\right)
   +C_6t^{\frac32+\xi}\,{}_2F_1\left(\frac{3}{4}+\frac{\xi}{2},
                \frac{5}{4}+\frac\xi2;1+\xi;-t^2\right)         \lab{6.6a}
\eea
where $\xi=\dis\frac{1}{2}\sqrt{1-8\mu^2\ell^2/\r}$ and ${}_2F_1(a,b;c;z)$
is the hypergeometric function \cite{x17}.\\
(d2) For $8\m^2\ell^2/\r=1$:
\bea
f_y=C_7t^{3/2}\,
   {}_2F_1\left(\frac{3}{4},\frac{5}{4};1;-t^2\right)
   +C_8G^{20}_{20}\left(-t^2\left|\ba{c} 1/2\,,1\\
                                   3/4\,,3/4
                                  \ea\right.\right)\, ,         \lab{6.6b}
\eea
\esubeq
where $G^{mn}_{pq}$ is the Meijer G function \cite{x17}. In both cases,
the associated solution for $f_x$ is given by $f_x=f_y/t$, see \eq{6.2},
and the metric function $h$ remains the same as in \eq{6.5}.

\begin{figure}[ht]
\centering
\includegraphics[height=4cm]{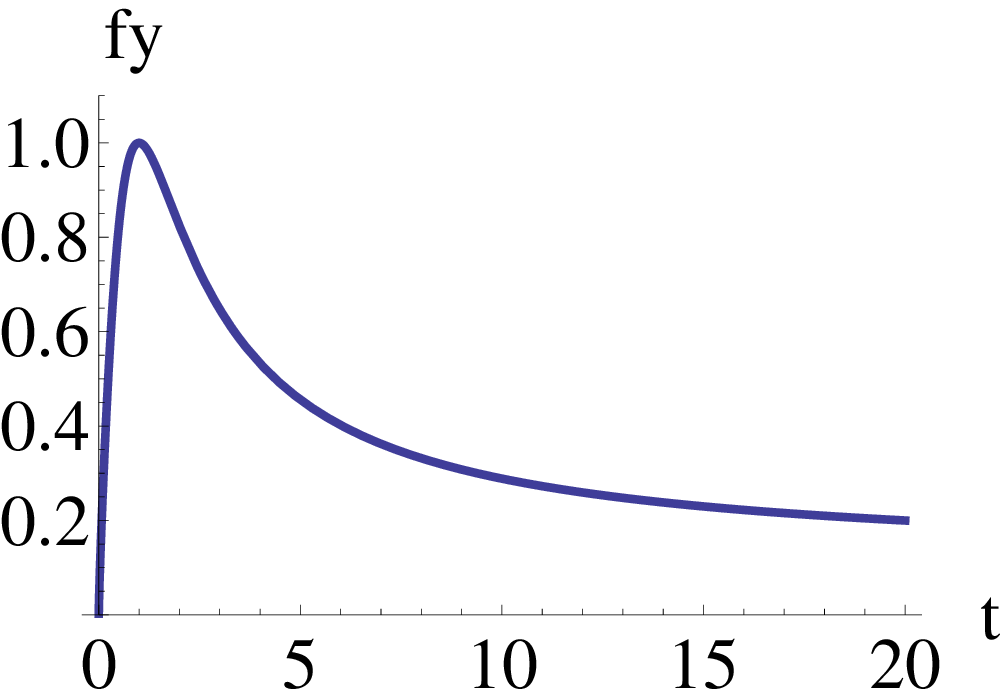}\qquad
\includegraphics[height=4cm]{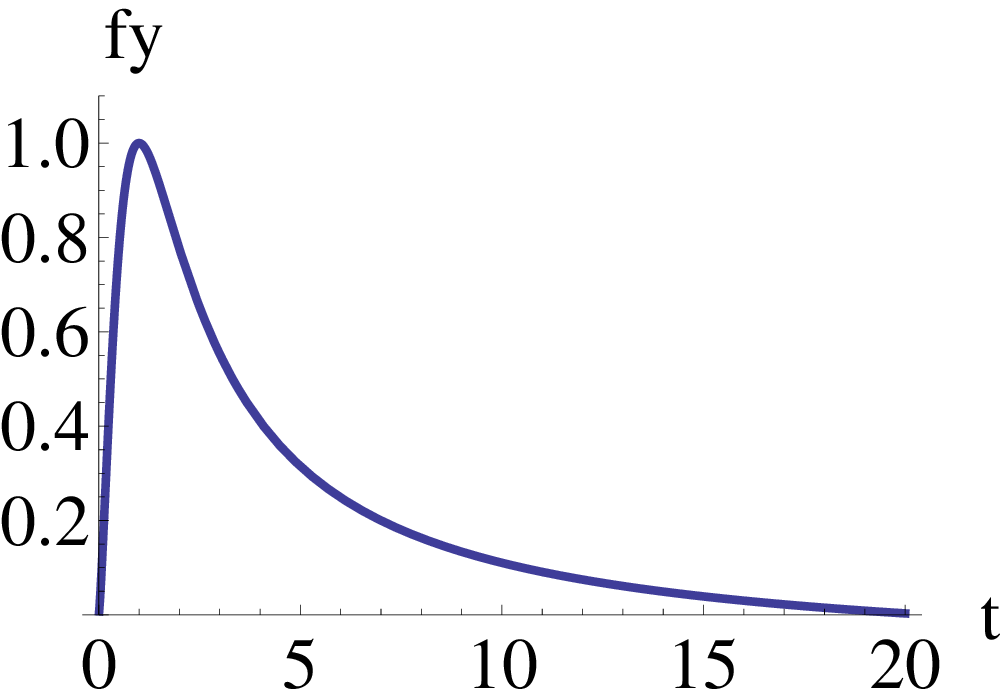}
\caption{The plots of the torsion function $f_y$ in \eq{6.6a}:
$8\m^2\ell^2/\r=-1$, $f_y[1]=1,f'_y[1]=0$  (left), and in \eq{6.6b}:
$f_y[1]=1,f'_y[1]=0$ (right).}\label{fig1}
\end{figure}
In the above two cases (d1) and (d2), the forms of the corresponding
torsion functions $f_y$ are illustrated in Figure \ref{fig1}.

\section{Exponential solution}
\setcounter{equation}{0}

In this section, we start with
\be
K_x=e^xf_x(y)\,,\qquad K_y=e^xf_y(y)\, ,\qquad H=e^xh(y)\, ,
\ee
whereupon the field equations \eq{4.5} become:
\bsubeq\lab{7.2}
\bea
&&(y^2 +\r y^2\pd_{yy}+2\m^2\ell^2)f_x
       +\bigl[(1-\r)y^2\pd_y-(1+\r)y\bigr]f_y=0\, ,             \lab{7.2a}\\
&&(y^2\pd_{yy} +\r y^2+2\m^2\ell^2)f_y
       +\bigl[(1-\r)y^2\pd_y+(1+\r)y\bigr]f_x=0\, ,             \lab{7.2b}\\
&&\hS H=\s(yf_x+y\pd_yf_y-f_y)\, ,
\eea
\esubeq
and $\hS H=e^x\left[y(h+h'')-2h'\right]$.

\begin{figure}[ht]
\centering
\includegraphics[height=4cm]{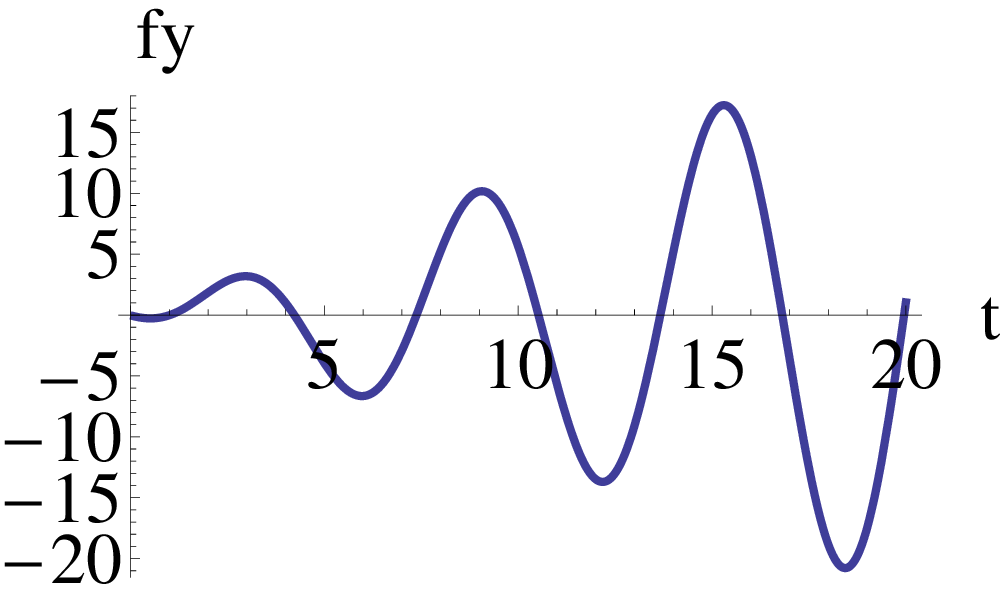}\qquad
\includegraphics[height=4cm]{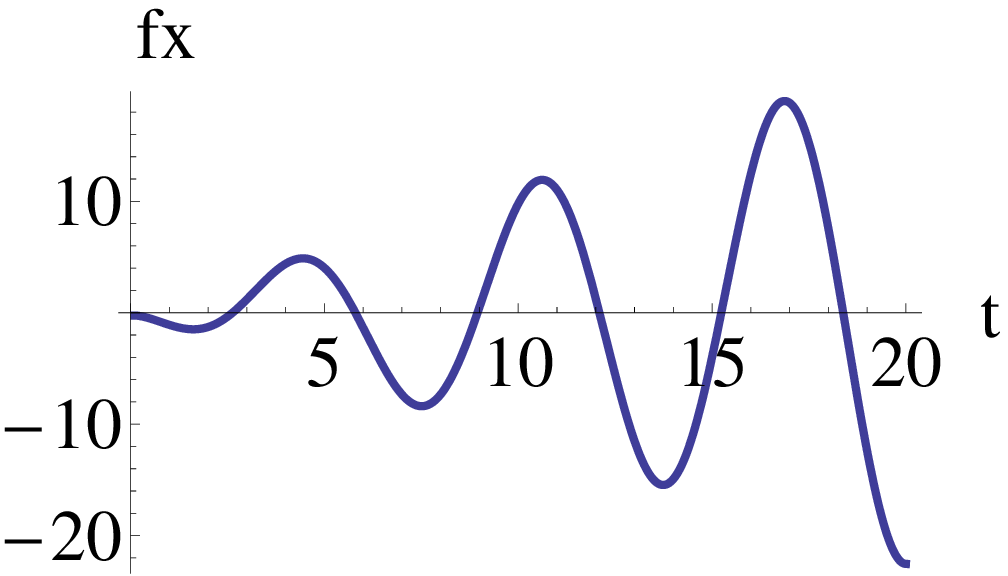}
\caption{The plots of the torsion functions \eq{7.7} for $D_1=D_2=1$,
$8\mu^2\ell^2/\r=-1$.}\label{fig2}
\end{figure}

As in the previous section, we assume that $H$ coincides with the vacuum
solution of \grl, defined by $\hS H=0$. This imposes an extra condition on
$f_x$ and $f_y$:
\be
yf_x+y\pd_yf_y-f_y=0\quad\Ra\quad
\frac{f_x}y+\left(\frac{f_y}y\right)'=0\, .                     \lab{7.3}
\ee
By introducing a change of variables, given by
\bsubeq\lab{7.4}
\be
f_x=yg_x\,,\quad f_y=yg_y\, ,                                   \lab{7.4a}
\ee
the condition \eq{7.3} takes a simple form:
\be
g_x+g'_y=0\, .
\ee
\esubeq
As a consequence, Eqs. \eq{7.2a} and \eq{7.2b} are transformed into
\bsubeq
\bea
&&\r y^2g_y^{(3)}+2\r yg_y''+(\r y^2+2\mu^2\ell^2)g_y'
                            +2\r yg_y=0\, ,                     \lab{7.5a}\\
&&\r y^2g_y''+(\r y^2+2\mu^2\ell^2)g_y=0\, .                    \lab{7.5b}
\eea
\esubeq
One can note that \eq{7.5a} is equal to the derivative (with respect to
$y$) of \eq{7.5b}. The solution of \eq{7.5b} reads:
\bea
g_y=\sqrt y\bigl[D_1J_\nu(y)+D_2Y_\nu(y)\bigr]\,,
\eea
where $\n=\frac{1}{2}\sqrt{1-8\mu^2\ell^2/\r}$, and $J_\nu$, $Y_\nu$ are
the Bessel functions of the first and second kind, respectively
\cite{x17}. Hence:
\bsubeq\lab{7.7}
\be
f_y=y^{\frac{3}{2}}\left(D_1J_\n(y)+D_2Y_\n(y)\right)\, ,
\ee
and $f_x=-yg'_y$ yields
\be
f_x=\sqrt{y}\Bigl[ D_1\Bigl(yJ_{\n+1}(y)-(\n+1/2)J_\n(y)\Bigr)
   +D_2\Bigl(yY_{\nu+1}(y)-(\n+1/2)Y_\n(y)\Bigr)\Bigr]\, .
\ee
\esubeq
The forms of the torsion functions \eq{7.7} are illustrated in Figure
\ref{fig2}.
They are of the same type as the \grl\ metric function $H_3$, defined in
section \ref{sec2}. Together, they define our third specific Siklos wave
with torsion.

\section{Concluding remarks}
\setcounter{equation}{0}

In this paper, we introduced a new class of exact vacuum solutions of PGT,
the Siklos waves with torsion. The solution is constructed in a way that
respects the radiation nature of the original Siklos configuration in
\grl. This is achieved by an ansatz for the RC connection that produces
only the tensorial irreducible mode of the torsion, propagating on the AdS
background. A compact form of the PGT field equations is used to find
three particular vacuum solutions belonging to the class of Siklos waves
with torsion; they generalize the Kaigorodov, the homogeneous and the
exponential solution of \grl.

\section*{Acknowledgments}

This work was supported by the Serbian Science Foundation under Grant No.
171031.

\appendix
\section{Irreducible decomposition of the field strengths}
\setcounter{equation}{0}

We present here formulas for the irreducible decomposition of torsion and
curvature in 4D Riemann--Cartan spacetime \cite{x5}; for general D, see
\cite{x19}.

It is convenient to start the exposition with the Bianchi identities:
\be
\nab T^i=R^i{_m}b^m\, ,\qquad \nab R^{ij}=0\, .                 \lab{A.1}
\ee

The torsion 2-form has three irreducible pieces:
\bea
&&{}^{(2)}T^i=\frac{1}{3}b^i\wedge(h_m\inn T^m)\, ,             \nn\\
&&{}^{(3)}T^i=-\frac{1}{3}\hd\bigl[b^i\wedge\hd(T^m\wedge b_m)\bigr]
             =\frac{1}{3}h^i\inn(T^m\wedge b_m)\, ,             \nn\\
&&{}^{(1)}T^i=T^i-{}^{(2)}T^i-{}^{(3)}T^i\, .
\eea

The RC curvature 2-form can be decomposed into six irreducible pieces:
$$
\ba{ll}
{}^{(2)}R^{ij}=-{}^*(b^{[i}\wedge\Psi^{j]})\, ,
           &{}^{(4)}R^{ij}=b^{[i}\wedge\Phi^{j]}\, ,            \\
{}^{(3)}R^{ij}=-\dis\frac{1}{12}X\,{}^*(b^i\wedge b^j)\, ,
           &{}^{(6)}R^{ij}=\dis\frac{1}{12}Wb^i\wedge b^j\, ,   \\
{}^{(5)}R^{ij}=\dis\frac{1}{2}b^{[i}\wedge h^{j]}\inn(b^m\wedge W_m)\,,
   \qquad  &{}^{(1)}R^{ij}=R^{ij}-\sum_{a=2}^6{}^{(a)}R^{ij}\, .
\ea
$$
where
\bea
&&W^i:=h_m\inn R^{mi}=\ric^i\, , \qquad W:=h_i\inn W^i=R\, ,    \nn\\
&&X^i:={}^*(R^{ki}\wedge b_k)\, ,\qquad X:=h_i\inn X^i\,.
\eea
and
\bea
&&\Phi_i:=W_i-\frac{1}{4}b_iW-\frac{1}{2}h_i\inn(b^m\wedge W_m)\,,\nn\\
&&\Psi_i:=X_i-\frac{1}{4}b_i X-\frac{1}{2}h_i\inn(b^m\wedge X_m)\, .
\eea
The trace and symmetry properties of ${}^{(n)}R_{ij}$ can be found in Ref.
\cite{x19}, p. 127. All these properties are satisfied by our ansatz.

For torsion-free solutions, the first Bianchi identity in \eq{A.1} implies
$X^i=0$, hence ${}^{(2)}R^{ij}$ and ${}^{(3)}R^{ij}$ vanish. Moreover,
$\ric_{[ij]}=0$ implies ${}^{(5)}R^{ij}=0$. The remaining three curvature
parts, 1-st, 4-th and 6-th, are the PGT analogues of the irreducible pieces
of the Riemannian curvature. In Riemannian geometry, ${}^{(1)}R^{ij}$
coincides with the Weyl (conformal) tensor,
$$
C^{ij}:=R^{ij}-\frac{1}{2}(b^i\ric^j-b^j\ric^i)+\frac{1}{6}Rb^ib^j\, ,
$$
but in the RC geometry, ${}^{(1)}R^{ij}$ differs from $C^{ij}$ by the
presence of torsion terms. Thus, ${}^{(1)}R^{ij}$ is a true extension of
$C^{ij}$ to the RC geometry. The 4-th component is defined in terms of the
symmetric traceless Ricci tensor,
\be
\Phi_i=\left(\ric_{(ij)}-\frac{1}{4}\eta_{ij}R\right)b^j\, .    \lab{A.5}
\ee
The above formulas are taken from Refs. \cite{x5,x19} with one
modification: the definition of $W^i$ is taken with an additional minus
sign (Landau--Lifshitz convention), and for consistency, the overall
signs of the 4th, 5th and 6th curvature parts are also changed.

\section{PGT field equations}
\setcounter{equation}{0}

The gravitational dynamics of PGT is determined by a Lagrangian
$L_G=L_G(b^i,T^i,R^{ij})$ (4-form), which is assumed to be at most
quadratic in the field strengths (quadratic PGT) and parity invariant. The
form of $L_G$ can be conveniently represented as
\be
L_G=-\hd(a_0R+2\L) +\frac{1}{2}T^iH_i+\frac{1}{4}R^{ij}H'_{ij}\,,\lab{B.1}
\ee
where $H_i:=\pd L_G/\pd T^i$ (the covariant momentum) and $H'_{ij}$ define
the quadratic terms in $L_G$:
\bsubeq\lab{B.2}
\be
H_i=2\sum_{n=1}^3\hd(a_n{}^{(n)}T_i)\, ,\qquad
H'_{ij}:=2\sum_{n=1}^6\hd(b_n{}^{(n)}R_{ij})\, .
\ee
Varying $L_G$ with respect to $b^i$ and $\om^{ij}$ yields the PGT field
equations in vacuum. After introducing the complete covariant momentum
$H_{ij}:=\pd L_G/\pd R^{ij}$ by
\be
H_{ij}=-2a_0\hd(b^ib^j)+H'_{ij}\, ,
\ee
\esubeq
these equations can be written in a compact form as
\bea
(1ST)&& \nab H_i+E_i=0\, ,                                      \nn\\
(2ND)&& \nab H_{ij}+E_{ij}=0\, ,                                \lab{B.3}
\eea
where $E_i$ and $E_{ij}$ are the gravitational energy-momentum and spin
currents:
\bea
&&E_i:=h_i\inn L_G-(h_i\inn T^m)H_m
                 -\frac{1}{2}(h_i\inn R^{mn})H_{mn}\, ,         \nn\\
&&E_{ij}:=-(b_iH_j-b_j H_i)\, .
\eea

The general field equations \eq{B.3} are used in section \ref{sec4} to
describe specific dynamical aspects of the Siklos waves with torsion. In
the Riemannian sector with $T^i=0$, we have $H_i=0$ and $E_{ij}=0$, and
the field equations \eq{B.3} reduce to the form given in section 3.

\end{document}